\documentclass[fleqn,usenatbib,useAMS]{mnras}
\usepackage[utf8x]{inputenc}
\usepackage[english]{babel}
\usepackage{times}
\usepackage{latexsym}
\usepackage{dsfont}
\usepackage{enumitem}

\RequirePackage{lineno}
\usepackage{graphicx}
\usepackage{amsmath}
\usepackage{amssymb}
\usepackage{ltxtable}

\usepackage{mathrsfs}
\usepackage[x11names]{xcolor}
\usepackage{natbib}
\usepackage{subfigure}
\usepackage{caption}
\usepackage{color}
\usepackage{epsfig}
\usepackage{multirow}

\newcommand{\gRays}{$\gamma$-rays}
\newcommand{\gray}{${\gamma}$-ray}
\newcommand{\xray}{\mbox{X-ray}}
\newcommand{\xrays}{\mbox{X-rays}}

\newcommand{\be}{\begin{equation}}
\newcommand{\ee}{\end{equation}}
\newcommand{\ba}{\begin{eqnarray}}
\newcommand{\ea}{\end{eqnarray}}

\newcommand{\pd}{\partial}  
\newcounter{ichi}
\newcounter{ni}
\newcounter{san}
\newcounter{yon}
\title[ALMA and NOEMA constraints on embryonic SLSN remnants]{ALMA and NOEMA constraints on synchrotron nebular emission from embryonic superluminous supernova remnants and radio--gamma-ray connection}

\author[K. Murase et al.]{
Kohta Murase$^{1,2,3,4}$, 
Conor M.~B. Omand$^{5}$,
Deanne L. Coppejans$^{6}$,
Hiroshi Nagai$^{7,8}$, 
\newauthor
Geoffrey C. Bower$^{9,10}$, 
Ryan Chornock$^{11}$,
Derek B. Fox$^{2,3}$,
Kazumi Kashiyama$^{5,12}$,
\newauthor
Casey Law$^{13}$, 
Raffaella Margutti$^{12}$, 
Peter M\'esz\'aros$^{1,2,3}$
\\
$^1$Department of Physics, The Pennsylvania State University, University Park, PA 16802, USA\\
$^2$Department of Astronomy \& Astrophysics, The Pennsylvania State University, University Park, PA 16802, USA\\
$^3$Center for Mulitmessenger Astrophysics, Institute for Gravitation and the Cosmos, The Pennsylvania State University, University Park, PA 16802, USA\\
$^4$Center for Gravitational Physics, Yukawa Institute for Theoretical Physics, Kyoto University, Sakyo-ku, Kyoto 606-8502, Japan\\
$^5$Department of Physics, Gradaute School of Science, University of Tokyo, Bunkyo-ku, Tokyo 113-0033, Japan \\
$^6$Center for Interdisciplinary Exploration and Research in Astrophysics (CIERA); Department of Physics \& Astronomy, Northwestern University, Evanston, IL 60208, USA\\
$^7$National Astronomical Observatory of Japan, Mitaka, Tokyo 181-8588, Japan\\
$^8$The Graduate University for Advanced Studies, SOKENDAI, Mitaka, Tokyo 181-8588, Japan\\
$^9$Academia Sinica Institute of Astronomy and Astrophysics, Hilo, HI 96720, USA\\
$^{10}$Department of Physics \& Astronomy, University of Hawai'i at M$\tilde{\rm a}$noa, Honolulu, HI 96822, USA\\
$^{11}$Astrophysical Institute; Department of Physics \& Astronomy, Ohio University, Athens, OH 45701, USA\\
$^{12}$Research Center for the Early Universe; Department of Physics, Graduate School of Science, University of Tokyo, Bunkyo-ku, Tokyo 113-0033, Japan\\
$^{13}$Cahill Center for Astronomy and Astrophysics, California Institute of Technology, Pasadena, CA 91125, USA\\
}

\date{Accepted XXX. Received YYY; in original form ZZZ}

\pubyear{2021}

\begin{document}
\label{firstpage}
\pagerange{\pageref{firstpage}--\pageref{lastpage}}
\maketitle

\begin{abstract}
Fast-rotating pulsars and magnetars have been suggested as the central engines of super-luminous supernovae (SLSNe) and fast radio bursts, and this scenario naturally predicts non-thermal synchrotron emission from their nascent pulsar wind nebulae (PWNe). We report results of high-frequency radio observations with ALMA and NOEMA for three SLSNe (SN 2015bn, SN 2016ard, and SN 2017egm), and present a detailed theoretical model to calculate non-thermal emission from PWNe with an age of $\sim1-3$~yr. We find that the ALMA data disfavors a PWN model motivated by the Crab nebula for SN 2015bn and SN 2017egm, and argue that this tension can be resolved if the nebular magnetization is very high or very low. Such models can be tested by future MeV-GeV gamma-ray telescopes such as AMEGO.
\end{abstract}

\begin{keywords}
supernovae --- fast radio bursts
\end{keywords}



\section{Introduction}
Recent observations have revealed the diversity among different types of core-collapse supernovae (SNe) and compact binary mergers. 
Super-luminous supernovae (SLSNe) are among the most luminous explosive phenomena, and their optical emission is likely to be powered by the central engine and/or interactions between the SN ejecta and circumstellar material~\citep[][for reviews]{MSC19,G19,C21}. In particular, the most popular explanation for Type I SLSNe (SLSNe-I) that are not accompanied by hydrogen signatures is the ``pulsar/magnetar-driven'' scenario, in which optical photons are radiated via thermalization of the rotation energy injected through pulsar winds~\citep[][]{ISJ+13,CWV+13,NSJ+13}. 
In general, the pulsar/magnetar engine is of interest in light of the diversity of transient phenomena~\citep[e.g.,][]{TCQ04,Z14,GJM+15,MMK+15,KMB+16,MMB+18}, since it potentially gives a unified view of SLSNe, stripped-envelope SNe, long gamma-ray bursts (GRBs), and even fast radio bursts (FRBs). 
It has also been invoked to explain some of the rapidly-rising optical transients~\citep{HKM17}.

It is known that pulsar wind nebulae (PWNe) such as the Crab nebula are efficient accelerators of electrons and positrons (leptons) and possibly also ions. 
Broadband non-thermal emission from the nebulae has been observed in many Galactic PWNe, suggesting that a significant fraction of the wind magnetic energy is used for particle acceleration around the termination shock~\citep[see a review,][]{GS06}. 
At early times, efficient thermalization of non-thermal photons occurs, and the observed SN light curves can be explained by adjusting the magnetic field $B$, initial period $P_i$, and ejecta mass $M_{\rm ej}$~\citep[e.g.,][]{Kasen_Bildsten_2010,Woosley_2010}. However, there is a large degeneracy in model parameters, and detecting high-energy signals is relevant for revealing not only the central engine but also particle acceleration at hidden environments~\citep[e.g.,][]{MMZ09,KPO13,MKK+15}. 

Searches for non-thermal signatures from nascent magnetars have been further motivated by FRB studies. 
If FRBs originate from young neutron stars or white dwarfs, they would be expected to occur inside nebulae. \cite{MKM16} proposed quasi-steady synchrotron emission as a probe of the FRB progenitors and their possible connection to pulsar/magnetar-driven SNe including SLSNe. 
Later, a persistent radio counterpart to FRB\,121102~\citep{2017ApJ...834L...7T} was reported~\citep[see reviews,][for recent developments]{PHL19,XWD21}. 
On the other hand, X-ray and radio observations of SLSNe have yielded upper limits \citep{MCM+18,BCM+18,CMG+18,LOK+19,EMO+21}, apart from SN\,2017ens~\citep{DC2017ens}, SN\,2020tcw~\citep{DC2020tcw,DM2020tcw} and PTF\,10hgi. The radio emission from PTF 10hgi may be explained by the nebular synchrotron emission~\citep{EBM+19,LOK+19,EMO+21}. 
Nascent nebular emission from binary neutron star mergers~\citep{YTK18,MTF+18,MBM19} and accretion-induced collapse~\citep{KM17,MBM19} has also been discussed.  

This work presents new results of ALMA and NOEMA observations in the $90-230$~GHz bands, together with a numerical model to describe the early non-thermal nebular emission and its simple analytical prescription. We demonstrate the power of such high-frequency radio data, given that the detectability at the GHz band is often limited by strong absorption. We suggest models that can avoid existing multi-wavelength constraints from radio to X-ray bands, pointing out the importance of the connection to soft gamma-rays. 

We assume cosmological parameters with $h=0.7$, $\Omega_m=0.3$ and $\Omega_\Lambda=0.7$. We also use notations as $Q_x=Q/10^x$ in the CGS unit except $t_{\rm yr}\equiv (t/1~{\rm yr})$ and $M_{\rm ej,1}=M_{\rm ej}/10M_\odot$.

%
\begin{figure}
\includegraphics[width=\linewidth]{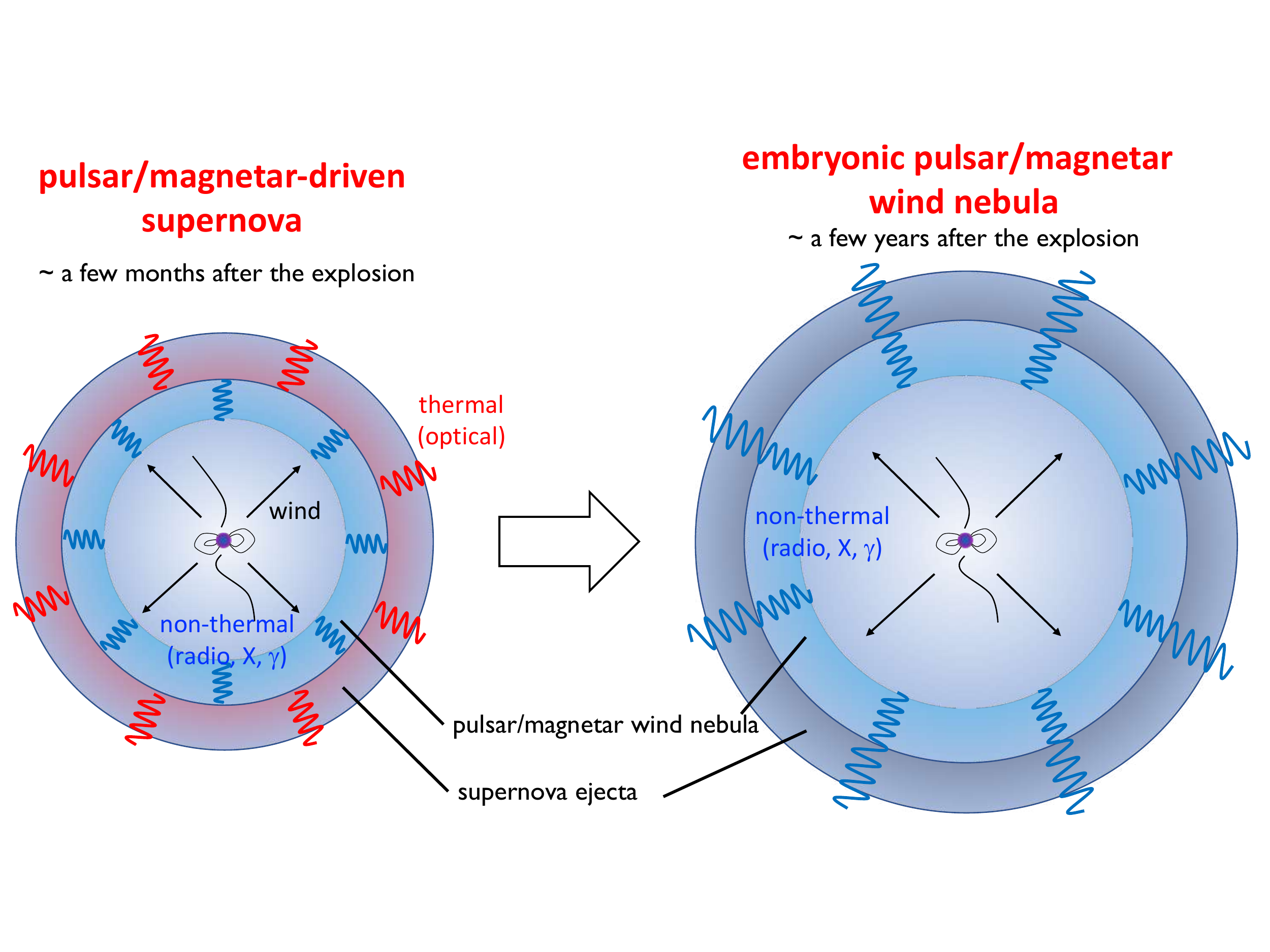}
\caption{Schematic picture of non-thermal radiation from embryonic SN remnants powered by pulsar/magnetar winds. Radio and gamma-ray emission is observable when the system is optically thin to various processes.  
}
\end{figure}
%

\section{Synchrotron emission from embryonic SLSN Remnants}
\subsection{Overview}
A highly magnetized pulsar/magnetar may be left over as a compact remnant after the SN explosion. 
Either rotational or magnetic energy is extracted from it by a relativistic wind, forming an embryonic PWN and powering the SN emission in the optical band via dissipation and thermalization of the magnetic energy. 
The spin-down power is estimated by~\citep{Gruzinov_2005,Spitkovsky_2006}
\begin{eqnarray}
L_{\rm sd}\approx7.2\times{10}^{41}~{\rm erg}~{\rm s}^{-1} \, B_{*,13}^2 \, P_{-2}^{-4},
\end{eqnarray}
where $B_*$ is the dipole magnetic field, $P(t)=P_i{(1+t/t_{\rm sd})}^{1/2}$ is the spin period, and the neutron star radius is assumed to be $12$~km.
The spin-down time is $t_{\rm sd}\approx0.12~{\rm yr}~B_{*,13}^2P_{i,-3}^{2}$.
An embryonic PWN is also expected to be a natural source of \gRays\ and hard \xrays\ (photon energies $E_\gamma=h\nu\gtrsim100$~keV range) with a long duration from months to years after the explosion~\citep{MKK+15,KMB+16}. 
In addition, $\sim1-10$~years after the explosion, quasi-steady synchrotron emission has been predicted to be detectable at the high-frequency radio band~\citep{OKM18}.

PWNe aid the expansion of the SN ejecta, and the nebular size $R_{\rm nb}$ is typically smaller than the ejecta radius $R_{\rm ej}$~\citep[e.g.,][]{CF92,S17}.
If the system is rotationally-powered, one expects $R_{\rm nb}\approx R_{\rm ej}\approx\beta_{\rm ej}ct\approx\sqrt{2{\mathcal E}_{\rm rot}/M_{\rm ej}}t\simeq5.2\times{10}^{16}~{\rm cm}~P_{i, -3}^{-1} M_{\rm ej,1}^{-1/2}t_{\rm yr}$ for $t>t_{\rm sd}$. 
Then, the magnetic field accumulated in the nebula is estimated to be $B_{\rm nb}\approx {(6\epsilon_B{\mathcal E}_{\rm rot}/R_{\rm nb}^3)}^{1/2}\simeq1.9~{\rm G}~\epsilon_{B,-2.5}^{1/2}P_{i,-3}^{1/2}M_{\rm ej,1}^{3/4}t_{\rm yr}^{-3/2}$, where $\epsilon_B\sim0.003$ is the energy fraction carried by nebular magnetic fields, as suggested for the Crab nebula~\citep[e.g.,][]{KC84mhd,TT10}.
Note that this value is highly uncertain in embryonic PWNe, and that we consider different magnetization cases in our treatment below.

The typical Lorentz factor of radio-emitting leptons is 
\begin{eqnarray}
\gamma(\nu)\simeq110~\epsilon_{B,-2.5}^{-1/4}P_{i,-3}^{-1/4}M_{\rm ej,1}^{-3/8}t_{\rm yr}^{3/4}\nu_{11}^{1/2}. 
\end{eqnarray}
The cooling Lorentz factor is given by $\gamma_c=6\pi m_ec/(\sigma_T B_{\rm nb}^2 t)\simeq6.8~\epsilon_{B,-2.5}^{-1}P_{i,-3}^{-1}M_{\rm ej,1}^{-3/2}t_{\rm yr}^{2}$. 
Thus, the synchrotron spectrum of embryonic PWNe is in the fast cooling regime, implying that the spectrum is softer than that of the Crab nebula with $\beta\sim1.3$ in the radio band. 
Here the photon index introduced by $F_\nu\propto\nu^{1-\beta}$ leads to $\beta={\rm max}[3/2,(2+q_1)/2]$, where $q_1<2$ is the low-energy spectral index of non-thermal leptons. 
The multi-wavelength modeling of young Galactic PWNe implies that the leptons are accelerated to $\gamma_b\in[3\times{10}^4,{10}^7]$~\citep{TT13}. 
The characteristic synchrotron frequency, at which $\nu F_\nu$ peaks, is 
\begin{eqnarray}
\nu_{b}\approx \frac{3}{4\pi}\gamma_b^2\frac{eB_{\rm nb}}{m_e c}\simeq3.2\times{10}^{18}~{\rm Hz}~\gamma_{b,5.8}^2\epsilon_{B,-2.5}^{1/2}P_{i,-3}^{1/2}M_{\rm ej,1}^{3/4}t_{\rm yr}^{-3/2},
\end{eqnarray}
given that this is lower than the maximum synchrotron frequency, $\nu_{M}\sim3.8\times{10}^{22}~{\rm Hz}$. 
Taking the reference frequency $\nu_0$ as the peak of $\nu F_\nu$ (i.e., $\nu_0=\rm max[\nu_b,\nu_M]$), $F_{\nu}$ in the fast cooling limit is 
\begin{eqnarray}
F_{\nu}=F_{\nu0}{\left(\frac{\nu}{\nu_0}\right)}^{1-\beta}\approx \frac{\epsilon_eL_{\rm sd}}{8\pi d^2\nu_0{\mathcal R}_0(1+Y_{\rm IC})}{\left(\frac{\nu}{\nu_0}\right)}^{1-\beta},
\end{eqnarray}
where $\epsilon_e\sim1$ is the energy fraction carried by the non-thermal leptons, ${\mathcal R}_0$ is a correction factor for the lepton normalization from the differential to bolometric powers, $Y_{\rm IC}$ is the Compton Y parameter, and $d$ is the distance to the source. 
In the case of $\beta=7/4$ (or $q_1=3/2$), we have $F_{\nu} \simeq 6.0~{\rm mJy}~\gamma_{b,5.8}^{-1/2}\epsilon_{B,-2.5}^{-1/8}\nu_{11}^{-3/4}B_{*,13}^{-2}P_{i,-3}^{-1/8}M_{\rm ej,1}^{-3/16}t_{\rm yr}^{-13/8}{(d/0.5~{\rm Gpc})}^{-2}$\\${[{\mathcal R}_0(1+Y_{\rm IC})/4]}^{-1}$, agreeing with numerical results (see below). 

Radio emission is subject to various absorption processes. In particular, synchrotron (self-)absorption (SSA) and free-free absorption are relevant. The SSA frequency $\nu_{\rm sa}$ can be estimated by
\begin{eqnarray}
\pi \frac{R_{\rm nb}^2}{d^2} 2 kT_{\rm sa} \frac{\nu_{\rm sa}^2}{c^2}= F_{\nu 0}{\left(\frac{\nu_{\rm sa}}{\nu_0}\right)}^{1-\beta},
\end{eqnarray}
where 
\begin{eqnarray}
T_{\rm sa}=\frac{1}{3k}C {\left(\frac{4\pi m_e c\nu_{\rm sa}}{3eB}\right)}^{1/2} m_e c^2
\end{eqnarray}
is the brightness temperature at $\nu_{\rm sa}$ and $C$ is an order-of-unity correction factor. We approximately have
\begin{eqnarray}
\nu_{\rm sa}\sim{\left(\frac{3^{3/2}e^{1/2}B_{\rm nb}^{1/2}F_{\nu_0}\nu_0^{\beta-1}d^2}{4\pi^{3/2}m_e^{3/2}c^{1/2}R_{\rm nb}^2} \right)}^{\frac{2}{2\beta+3}},
\end{eqnarray}
which leads to $\nu_{\rm sa}\sim32~{\rm GHz}~\epsilon_{B,-2.5}^{1/26}\gamma_{b,5.8}^{-2/13}B_{*,13}^{-8/13}P_{i,-3}^{17/26}M_{\rm ej,1}^{19/52}$\\$t_{\rm yr}^{-35/26}{[{\mathcal R}_0(1+Y_{\rm IC})/4]}^{-4/13}$ for $\beta=7/4$, agreeing with numerical results presented below. 
In reality, non-thermal emission escaping from the nebula is further degraded by free-free absorption in the SN ejecta, and the nebula typically becomes transparent at the 100~GHz band $\sim$1--10~years after the explosion~\citep{MKM16,OKM18,MMB+18}.

\subsection{Theoretical Model}
We here outline the theoretical model used for calculations of thermal and non-thermal spectra. 
Although method is used is similar to that of \cite{OKM18} and \cite{EMO+21}, we review it so that we may provide details not presented in these works. 
Also, there are several improvements compared to \cite{MKM16}. 
In the pulsar/magnetar-powered SN scenario, a significant fraction of the spin-down energy needs to be deposited in the SN ejecta, and the evolution of the internal energy ${\mathcal E}_{\rm int}$ is given by
\begin{eqnarray}
\frac{d{\mathcal E}_{\rm int}}{dt}=f_{\rm dep,\rm sd}\epsilon_eL_{\rm sd}+f_{\rm dep, \rm rd}L_{\rm rd}-L_{\rm sn}-\frac{{\mathcal E}_{\rm int}}{R_{\rm ej}}\frac{dR_{\rm ej}}{dt}
\end{eqnarray}
where $L_{\rm rd}$ is the radioactive decay power, $f_{\rm dep,\rm sd/rd}$ is the energy fraction deposited into thermal energy, $L_{\rm sn}$ is the SN luminosity, and the last term represents adiabatic losses. We estimate $f_{\rm dep,\rm sd}$ as in \cite{KMB+16}, but the treatment is improved by considering arbitrary $\gamma_b$ and $q_1<2$ rather than assuming $q_1=2$. 

Dynamics of PWNe and SN ejecta can be calculated by solving equations of motion for the shocked shells~\citep{OG71}. \cite{MKM16} studied radio and millimeter emission from nascent magnetars using analytical solutions for the PWN evolution. Instead, we estimate radii of the nebula and ejecta by solving the simplified equations~\citep{MVH+14,KMB+16},
\begin{eqnarray}
\frac{dR_{\rm nb}}{dt}&=&V'_{\rm nb}+\frac{R_{\rm nb}}{t},\\\nonumber
\frac{dR_{\rm ej}}{dt}&=&V_{\rm ej},
\end{eqnarray}
where $V'_{\rm nb}$ is the nebular velocity in the ejecta rest frame and $V_{\rm ej}$ is the ejecta velocity, respectively. For engine-powered SLSNe, the ejecta and nebula move together, i.e., $R_{\rm ej}\approx R_{\rm nb}$. 
The evolution of nebular magnetic fields is given by
\begin{eqnarray}
\frac{d{\mathcal E}_{B}}{dt}=\epsilon_B L_{\rm sd}-c_B\frac{{\mathcal E}_{B}}{R_{\rm nb}}\frac{dR_{\rm nb}}{dt},
\end{eqnarray}
where ${\mathcal E}_{B}$ is the magnetic energy inside the nebula. The magnetic field is uncertain and the toroidal component may be accumulated in the nebula. In this work, we consider the limit $c_B=0$, as used in the modeling of Galactic PWNe~\citep{TT10} and our past works~\citep{MKK+15,OKM18}. 

We calculate intrinsic non-thermal emission from the nebula by solving the following kinetic equations, 
\begin{eqnarray}\label{eq:cascade}
\dot{n}_{E_e}^e &=& \dot{n}_{E_e}^{(\gamma \gamma)}
- \frac{\pd}{\pd E} [(P_{\rm IC}+P_{\rm syn}+P_{\rm ad}) n_{E_e}^e] 
+ \dot{n}_{E_e}^{\rm inj},\nonumber\\
\dot{n}_{E_\gamma}^\gamma &=& -\frac{n_{E_\gamma}^{\gamma}}{t_{\gamma \gamma}} - \frac{n_{E_\gamma}^{\gamma}}{t_{\rm comp}^{\rm nb}}- \frac{n_{E_\gamma}^{\gamma}}{t_{\rm esc}^{\rm nb}}
+ \dot{n}_{E_\gamma}^{(\rm IC)} 
+ \dot{n}_{E_\gamma}^{(\rm syn)},
\end{eqnarray}
where $t_{\gamma\gamma}$ is the two-photon annihilation time, $t_{\rm comp}^{\rm nb}$ is the energy-loss time due to Compton scatterings in the nebula, $t_{\rm esc}^{\rm nb}=R_{\rm nb}/c$ is the photon escape time for non-thermal photons, and $P_{\rm IC}$, $P_{\rm syn}$ and $P_{\rm ad}$ are energy-loss rates due to the IC, synchrotron radiation and adiabatic expansion, respectively. 
Note that different from \cite{MKK+15} and \cite{MKM16}, we take into account both electromagnetic cascades and contributions from relic pairs that are injected at $t<t_{\rm sd}$. These pairs are relevant at $t>t_{\rm sd}$ if $q_1\lesssim1$ .   
The lepton injection rate $\dot{n}_{E_e}^{\rm inj}$ is determined by 
\begin{eqnarray}
E_e^2\dot{n}_{E_e}^{\rm inj}=\frac{3\epsilon_eL_{\rm sd}}{4\pi R_{\rm nb}^2 c{\mathcal R}_0} 
\left\{ \begin{array}{ll} 
{(\gamma_e/\gamma_b)}^{2-q_1}
& \mbox{($\varepsilon \leq \varepsilon_b$)}\\
{(\gamma_e/\gamma_b)}^{2-q_2}
& \mbox{($\varepsilon_b < \varepsilon$)}
\end{array} \right.
\label{eq:inj}
\end{eqnarray}
where $q_1<2$ and $q_2\geq2$ are injection spectral indices. 
The observations of known Galactic PWNe suggest that a significant fraction of the spin-down power is dissipated inside or around the termination shock~\citep[e.g.,][]{TT13}, and we take $\epsilon_e=1-\epsilon_B$. 
Note that efficient conversion from the rotation energy to the particle energy is also necessary to explain the observed optical emission in the pulsar/magnetar-driven scenario. 
We treat $\gamma_b$ as a parameter, assuming $\gamma_e\in[10^3,10^7]$, where $\gamma_b$ does not have to be the same as the bulk Lorentz factor of the wind. 
The pair multiplicity is model dependent, and cascades in the nebula and/or wind may contribute at early times~\citep{MKK+15,VM21}. 
Also, it has been known that the radio data of the Crab nebula require large multiplicities (with $\sim10^6$) that are theoretically challenging~\citep[e.g.,][]{A12}, which may originate from continuously-heated pairs that were injected in the past or from another component~\citep{A99,TA17}. 
Thus, we use Eq.~(\ref{eq:inj}) to allow such possibilities, in which the effective pair multiplicity can be expressed by $\gamma_b$, $q_1$ and $q_2$~\citep{MKK+15}. 
Leptons above $\gamma_b$ may be accelerated in shocks, but will not directly affect the radio flux. 
We adopt $q_2=2.5$ throughout this work, and note that this choice does not significantly affect our results on radio fluxes. 

With respect to absorption processes, we implement free-free absorption and the Razin-Tsytovich suppression in the ejecta as well as SSA in the nebula. 
Following \cite{MKM16}, we consider two representative cases with and without absorption in the singly-ionized CO ejecta. 
In reality, clumpiness or asymmetry in the ejecta leads to low-density regions through which the radio emission can escape more easily~\citep[e.g.,][]{SM21}. 
Note that the recent detection of radio emission from PTF10hgi could be explained by quasi-steady emission from nascent PWNe, for which the required absorption coefficient lies within the two limits~\citep{EBM+19,LOK+19,EMO+21,HTM+21}. 
 

\begin{table*}
\caption{Model parameters used for fitting optical light curves of three SLSNe. Initial periods are investigated from 1.0 ms to $P_{\text{max}}$. For $P_i=1$~ms, the high magnetization model is also considered (see text for details), for which the parameters are indicated in parentheses.
}
\begin{center}
\begin{tabular}{|c||c|c||c|c||c|c|c|} \hline
SN name & $z$ & $T_{\rm pk}$ [MJD] & $B_{*,13}$ at $P=1$~ms & $M_{\rm ej}$ ($M_{\sun}$) at $P=1$~ms & $P_{\text{max}}$ (ms) &  $B_{*,13}$ at $P_{\text{max}}$ & $M_{\rm ej}$ ($M_{\sun}$) at $P_{\text{max}}$ \\ \hline 
2015bn & 0.1136 & 57102  & 2.1 (1.0) & 17 (8.5) & 1.4 & 1.0 & 5.0 \\
2016ard & 0.2025 & 57463  & 6.0 (3.0) & 12 (6.0) & 2.2 & 1.7 & 1.5 \\
2017egm & 0.03072 & 57922 & 13 (6.5) & 11.5 (5.7) & 2.0 & 2.0 & 2.0 \\ \hline
\end{tabular}
\end{center}
\label{tab:SLSNe}
\end{table*}
\begin{table}
\centering
\caption{Summary of ALMA (upper columns) and NOEMA (lower columns) observations of the three SLSNe. The former was based on the project codes 2017.1.00975.S (PI: K. Murase) and 2018.1.01295.S (PI: D. Coppejans), and the latter was based on the project ID S18BH (PI: C. Omand). Note that $T=t+T_i$. }
\begin{tabular}{lcccc}
\hline
SN name & $T$ & Integration time & Frequency & Upper limit \\
 & [MJD] & [s] & [Hz] & [$\mu$Jy] (3$\sigma$) \\ \hline
2015bn & 58218 & 333 & 97.1 & 87.4 \\
 & 58216 & 666 & 233.0 & 123 \\
2016ard & 58317-58318 & 5746 & 97.5 & 26.0 \\
 & 58292-58297 & 28123 & 233.0 & 15.1 \\
2017egm & 58591 & 1182 & 97.5 & 134 \\
 & 58587 & 2331 & 233.0 & 151 \\\hline
2017egm & 58342 & 6800 & 86.3/101.7 &  \\
 & 58349 & 9400 & 86.3/101.7 & \\
 & 58368 & 9400 & 86.3/101.7 & \\
 & 58381 & 4000 & 86.3/101.7 & \\
 & 58382 & 4000 & 86.3/101.7 & \\
 & 58385 & 6800 & 86.3/101.7 & \\
(total) & 58342-58385 & 40300 & 86.3 & 40.4 \\
 & (58365: avg) & 40300 & 101.7 & 43.5 \\ \hline
\end{tabular}
\label{tab:obs}
\end{table} 

\section{Implications of High-Frequency Radio Data}
\subsection{ALMA and NOEMA observations}
Working from the Open Supernovae Catalog~\citep{OpenSN17}, \citet{OKM18} investigated the detectability of radio and millimetre emission from the recent brightest 6 SLSNe with good quality data of SN light curves. 
We supplement those events with four SLSNe that occurred over 2015--6, extracting the estimated spin-down parameters, $B_{*}$ and $P_i$, for each (see Table~\ref{tab:SLSNe}). 
Among these 10 SLSNe, we found just two objects, SN\,2015bn and SN\,2016ard, to be promising mm targets for ALMA. 
Subsequently, the nearby SN\,2017egm was discovered and added to our sample, with observations by ALMA and NOEMA conducted. 
The ALMA data are analyzed with the Common Astronomy Software Application (CASA) software package~\citep{CASA07}. 
The data were taken in the Time Division Mode at ALMA Bands 3 and 6. The bandwidths were 7.5~GHz, and two polarization products XX and YY were obtained to produce the Stokes I image.
The NOEMA data were analyzed with the Grenoble Image and Line Data Analysis Software (GILDAS)\footnote{http://www.iram.fr/IRAMFR/GILDAS} package. 
The data were taken in dual-polarization at NOEMA band 1, which includes two side bands centred at 86.26 GHz and 101.74 GHz with 16 GHz bandwidth each, giving the total coverage of 32 GHz.
With standard imaging techniques used in CASA and GILDAS, we obtain flux densities and root-mean-square values in the image within $5^{\prime\prime}$, and 3$\sigma$ upper limits are reported in Table~\ref{tab:obs}. 
We have confirmed that the sensitivity levels achieved by these observations are consistent with expectations, given the noise and integration times. 

{\bf SN 2015bn:} 
SN\,2015bn is among the best studied SLSNe-I~\citep[e.g.,][]{NBS+16,NBM+16}. It was discovered by the Catalina Sky Survey on 2014 December 23, and later observed by the Mount Lemmon Survey and the Pan-STARRS Survey. 
The sky coordinates are ${\rm RA}=11^{\rm h}33^{m}41^s$ and $\delta=+00^\circ43^\prime32^{\prime\prime}$ (J2000.0), and the measured redshift is $z=0.1136$. 
The optical light curve reached its peak on MJD 57102, which is $116$~days after the explosion time $T_i$ in our model. 
We adopt the values of $B_{*}$, $P_i$ and $M_{\rm ej}$ obtained by \cite{OKM18} with the V band data. ALMA observations at Band 3 ($\sim3$~mm) and Band 6 ($\sim1.3$~mm) were conducted on 2017 April 10 and April 8, respectively, about $1100$~days after the peak time $T_{\rm pk}$. The numbers of antennas were 46 and 43, respectively, for which 1 Execution Block (EB) was taken. 
The beam sizes were 1.87" by 1.55" at Band 3 and 1.02" by 0.64" at Band 6, respectively. 
The data were calibrated and imaged by the ALMA Pipeline version ver40896 with the CASA version 5.1.1-5.
No significant emission was found. 
Note that these observations are independent of those reported by \cite{EMO+21} at 100~GHz. 

{\bf SN 2016ard:}
SN\,2016ard was found by Pan-STARRS1 on 2018 February 14~\citep{CBK+16}. 
Its sky coordinates are ${\rm RA}=14^{\rm h}10^{m}44^s$ and $\delta=-10^\circ09^\prime35^{\prime\prime}$ (J2000.0), and its redshift is $z=0.2025$. 
The SN light curve in the $w$ band reached the maximum on MJD 57463, which is $74$~days after the explosion in our model. 
ALMA data at Band~3 and Band~6 were taken on 2018 July 18-19 and June 23-28, respectively, about 900~days after $T_{\rm pk}$. 
Correspondingly, 2 EBs and 6 EBs were obtained with 45 antennas and 45-47 antennas, respectively. 
The beam sizes were 3.17" by 2.23" at Band 3 and 1.27" by 0.94" at Band 6, respectively.
For each band, we used the ALMA Pipeline version ver40896 with the CASA version 5.1.1-5 and the ALMA Pipeline version ver42030M with the CASA version 5.4.0-68. 
Although the 100~GHz image had a $\sim1\sigma$ fluctuation, no significant emission was found. 

{\bf SN 2017egm:}
SN\,2017egm was discovered by the {\it Gaia} satellite on 2017 May 23~\citep{BSP+18}. 
Its sky coordinates are ${\rm RA}=10^{\rm h}19^{\rm m}05^{\rm s}$ and $\delta=+46^\circ27^\prime14^{\prime\prime}$ (J2000.0), and its redshift is $z=0.03072$. 
This SLSN-I, associated with the massive spiral galaxy NGC~3231, is one of the nearest SLSNe. The SN light curve reached its optical maximum on MJD 57463, which is $51$~days after the explosion in our model. 
The parameters, $B_{*}$, $P_i$ and $M_{\rm ej}$, are determined using the V band data. 
NOEMA data were taken during 6 epochs between 2018 August 12 and September 24 and calibrated and imaged using GILDAS version nov18a to obtain time-integrated, polarization-averaged upper limits at 86~GHz and 102~GHz. The corresponding beam sizes were 4.99" by 4.92" and 4.41" by 4.30", respectively. Emission was detected from the host galaxy at an offset of $13^{\prime\prime}$ from the SN position, consistent with the location of a known star-forming region \citep{2017ApJ...845L...8N}.
ALMA observations at Band 3 and Band 6 were conducted on 2019 April 18 and April 14, respectively. The numbers of antennas were 48 and 43, respectively, for which 1 EB was taken. 
The data were calibrated and imaged by the ALMA Pipeline version ver42254M with the CASA version 5.4.0-70. 
The beam sizes were 2.69" by 0.92" at Band 3 and 1.17" by 0.35" at Band 6, respectively.

\subsection{Model implications}
%
\begin{figure*}
\includegraphics[width=0.32\linewidth]{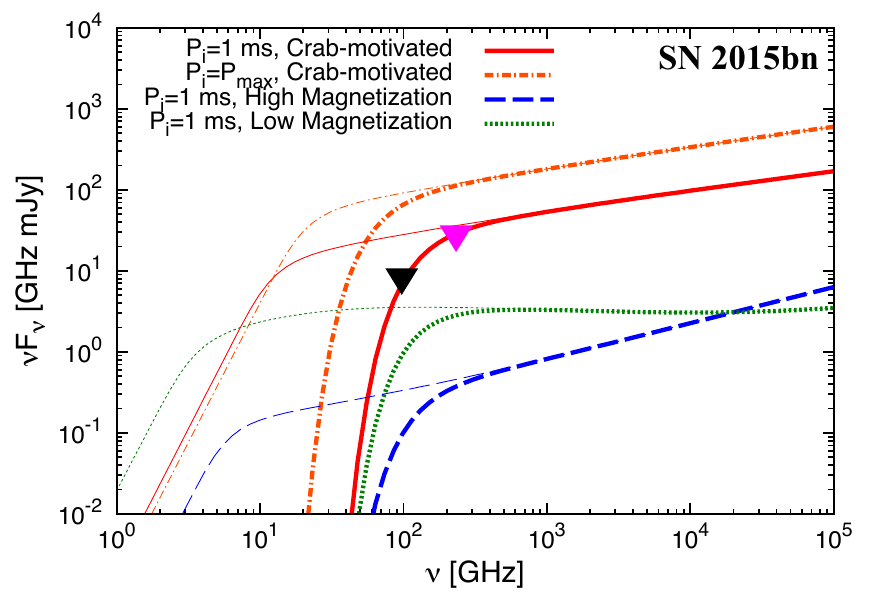}\hfill
\includegraphics[width=0.32\linewidth]{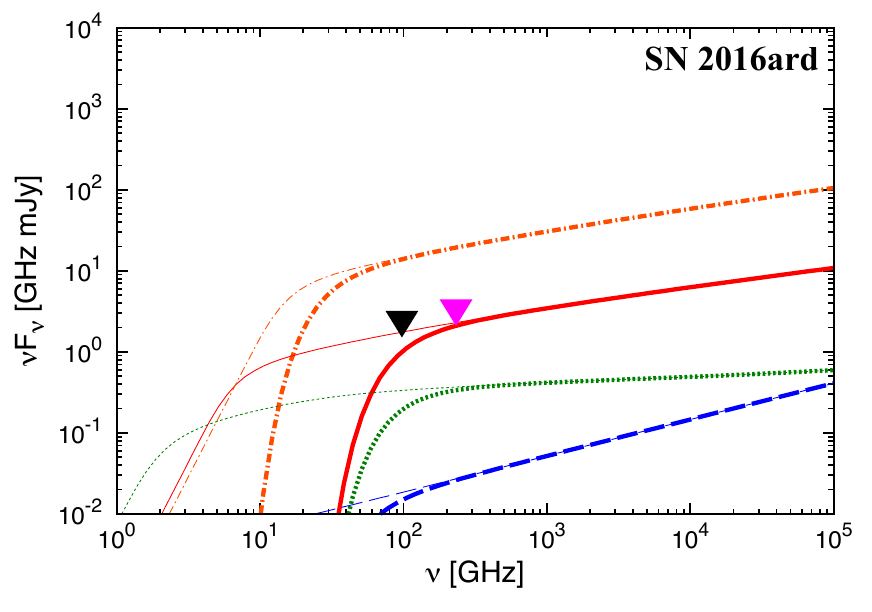}\hfill
\includegraphics[width=0.32\linewidth]{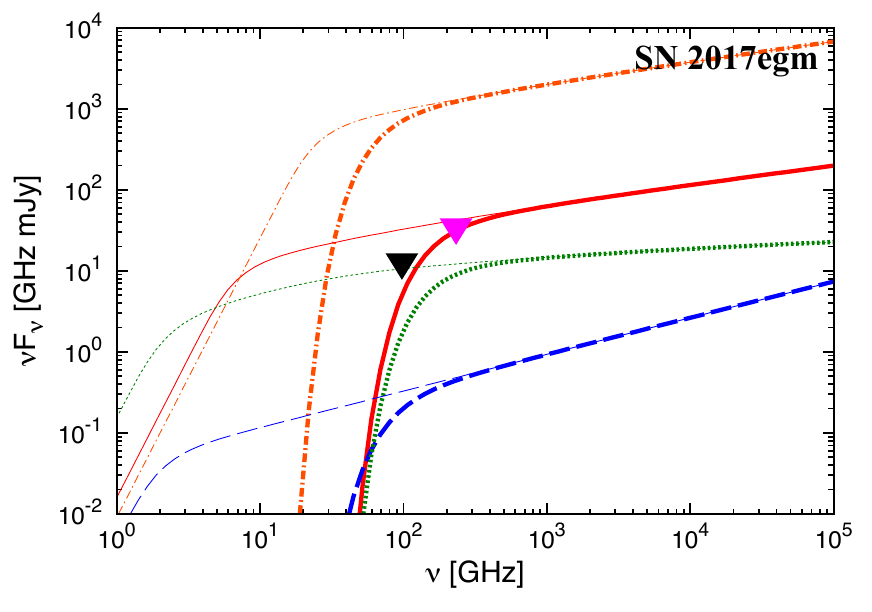}
\caption{Non-thermal spectra of SN 2015bn (left), SN 2016ard (middle), and SN 2017egm (right), at ALMA observation times for Band 3. Thick/dashed curves indicate synchrotron fluxes with/without ejecta absorption.
}
\label{fig:SP}
\end{figure*}
%
\begin{figure*}
\includegraphics[width=0.32\linewidth]{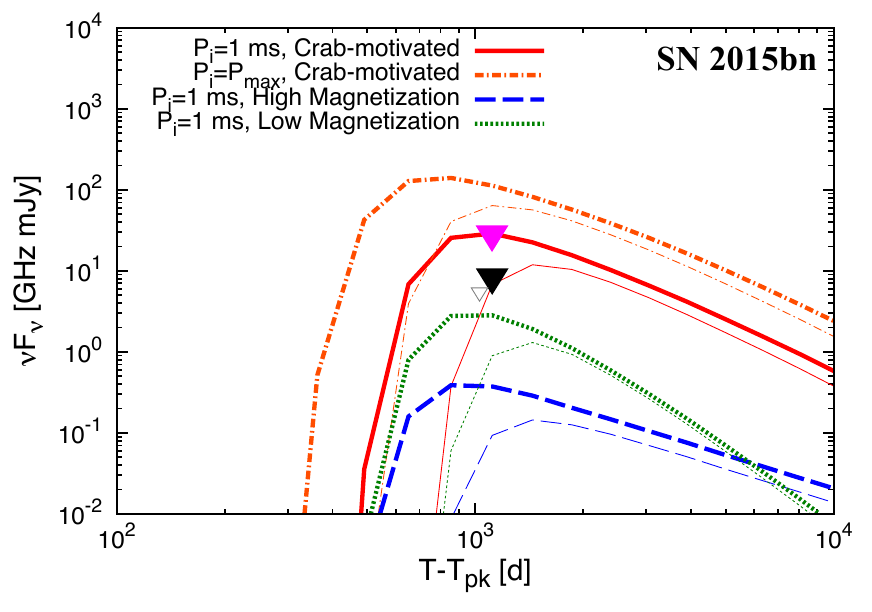}\hfill
\includegraphics[width=0.32\linewidth]{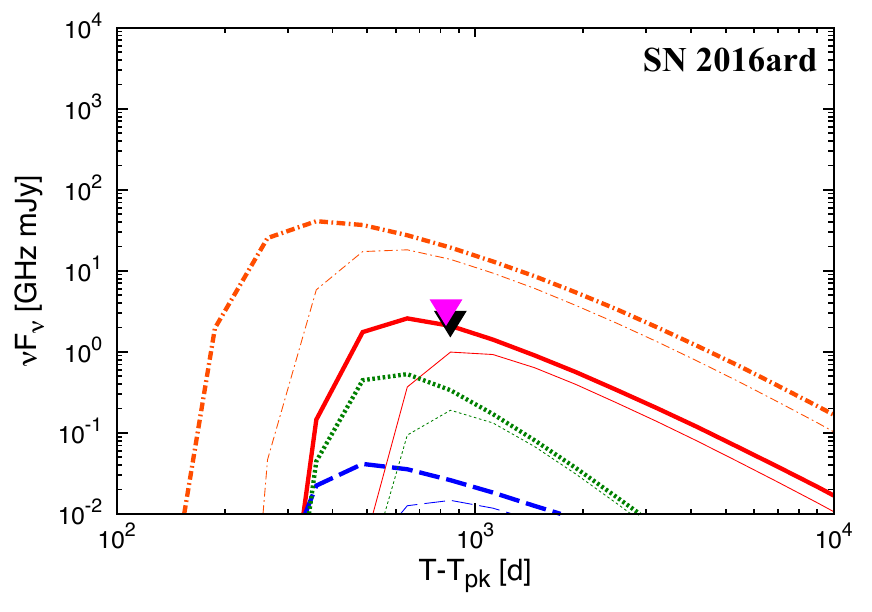}\hfill
\includegraphics[width=0.32\linewidth]{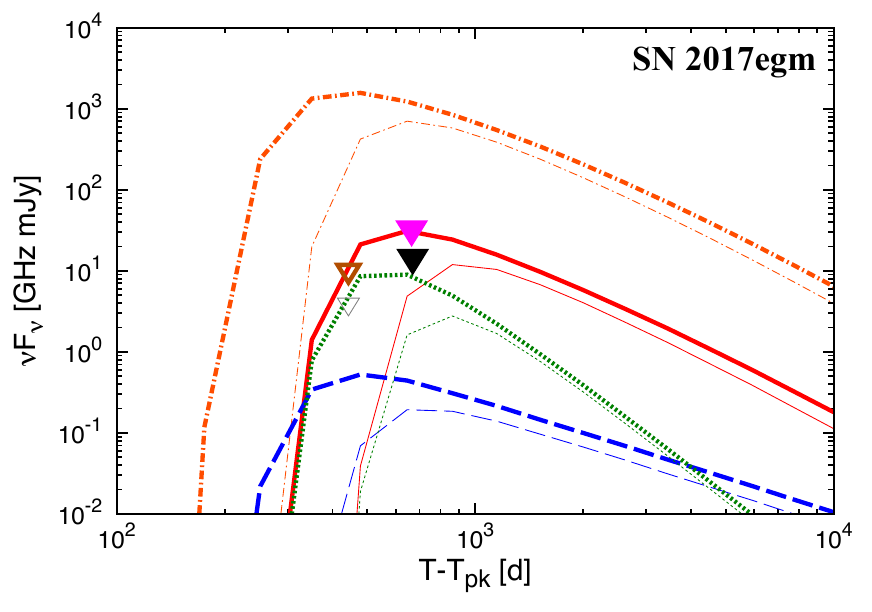}
\caption{High-frequency radio light curves of SN 2015bn (left), SN 2016ard (middle), and SN 2017egm (right) at two frequency bands.
Thick/thin curves represent model light curves at the 230~GHz/100~GHz band. Correspondingly, ALMA upper limits at Band 3 and Band 6 are shown as filled black and pink triangles, respectively. In the right panel, thick-brown/thin-gray open triangles in the right panel indicate NOEMA upper limits at the 100~GHz/90~GHz band. The gray open triangle in the left panel indicates the upper limit reported by \citet{EMO+21}.
}
\label{fig:LC}
\end{figure*}
%

Using the method described in Sec.~2.2, we calculate non-thermal PWN spectra in a time-dependent manner. 
The evolution of nascent PWNe is mainly governed by $B_*$, $P_i$ and $M_{\rm ej}$, which are determined through modeling SN optical light curves. 
The optical opacity, nickel mass, and initial ejecta energy are fixed to $K=0.1~{\rm cm}^2~{\rm g}^{-1}$, $M_{\rm Ni}=0.1~M_\odot$, and ${\mathcal E}_{\rm ej}=10^{51}$~erg, respectively. 
For non-thermal emission in the radio and millimetre bands, 3 microphysical parameters ($\epsilon_B$, $\gamma_b$, $q_1$) are relevant, for which we examine the following 3 models. 

{\bf Crab-motivated model:}
As the fiducial assumption, following \cite{MKK+15}, we postulate that the magnetization and lepton injection spectrum are similar to those inferred from young Galactic PWNe. 
In particular, motivated by the Crab nebula, we take $\epsilon_B=0.003$, $\gamma_b=6\times10^5$, and $q_1=1.5$~\citep[][]{TT10}. 
Note that $q_1>1$ is obtained by other studies \citep[e.g.,][]{AA96} and favored by modeling of radio emission from PTF\,10hgi~\citep{LOK+19}.  
Results are shown in Figs.~\ref{fig:SP} and \ref{fig:LC}. 
Because electrons and positrons are in the fast cooling regime, i.e., $\gamma_c<\gamma_b$, the resulting synchrotron spectra should be softer than those of Galactic PWNe. 
For SN\,2015bn and SN\,2017egm, light curves at 100~GHz and 230~GHz have peaks at $\sim600-1000$~d, depending on details of the free-free absorption in the ejecta.
With $P_i=1$~ms, the flux predictions for SN 2016ard and SN 2017egm are allowed by the data, while the models with $P_i=P_{\rm max}$ are ruled out. 
On the other hand, we find that the model fluxes of SN 2015bn and SN 2017egm have $\sim(2-3)\sigma$ tensions with the non-detections at 230~GHz even for $P_i=1$~ms. 
In this case, using $F_\nu\lesssim100~\mu$Jy at this frequency (see Table~\ref{tab:obs}), we have the following constraint,
\begin{eqnarray}
\gamma_{b,5.8}^{-1/2}\epsilon_{B,-2.5}^{-1/8}{(1+Y_{\rm IC})}^{-1}\lesssim1.
\end{eqnarray}
One sees that this requirement can be relaxed by increasing $\gamma_b$ and/or $\epsilon_B$ and/or $Y_{\rm IC}$, and we consider two alternative models.  

{\bf Low-magnetization model:} 
One of the solutions to reduce the radio flux is to increase $Y_{\rm IC}$. 
\cite{MKK+15} showed that external inverse-Compton emission is important until SN photons mostly escape. 
Intriguingly, extremely small values of $\epsilon_B$ are independently motivated by a possible solution to the missing energy problem for SN\,2015bn and SN\,2017egm~\citep{VM21}. 
As an example, we take $\epsilon_B={10}^{-6}$, $\gamma_b=10^3$, and $q_1=1$ (in which the Compton parameter $Y_{\rm IC}\gtrsim10$). 
Results are shown in Figs.~\ref{fig:SP} and \ref{fig:LC}, where we see that the synchrotron fluxes can be lower by an order of magnitude. 
Note that the spectrum is softer because the synchrotron peak is lower. 

{\bf High-magnetization model:} 
Here we consider an alternative model to satisfy the ALMA and NOEMA constraints, in which we adopt $\epsilon_B=0.5$, $\gamma_b=10^7$, and $q_1=1$. 
The magnetization around the termination shock may be as large as $\gtrsim1$. Although too large values cannot explain optical SN emission, dissipation and thermalization of the Poynting energy in the pulsar wind may be more inefficient. Such high-magnetization models inevitably give a stronger magnetic field in the nebula, $B_{\rm nb}\simeq24~{\rm G}~\epsilon_{B,-0.3}^{1/2}P_{i,-3}^{1/2}M_{\rm ej,1}^{3/4}t_{\rm yr}^{-3/2}$, and the synchrotron frequency, $\nu_{b}\simeq1.0\times{10}^{22}~{\rm Hz}~\gamma_{b,7}^2\epsilon_{B,-0.3}^{1/2}P_{i,-3}^{1/2}M_{\rm ej,1}^{3/4}t_{\rm yr}^{-3/2}$, is expected in the MeV range. 
Note that the fast cooling nebular spectrum is flatter than that observed in the Crab nebula~\citep{MKM16}. 
In this model, we have $F_{\nu} \simeq5.7~{\mu\rm Jy}~\gamma_{b,7}^{-1}\epsilon_{B,-0.3}^{-1/4}\nu_{11}^{-1/2}B_{*,13}^{-2}P_{i,-3}^{-1/4}M_{\rm ej,1}^{-3/8}t_{\rm yr}^{-5/4}{[{\mathcal R}_0(1+Y_{\rm IC})]}^{-1}$\\${(d/0.5~{\rm Gpc})}^{-2}$ and $\nu_{\rm sa}\sim4.3~{\rm GHz}~\gamma_{b,7}^{-1/3}B_{*,13}^{-2/3}P_{i,-3}^{2/3}M_{\rm ej,1}^{1/3}t_{\rm yr}^{-4/3}$\\${[{\mathcal R}_0(1+Y_{\rm IC})]}^{-1/3}$, consistent with our numerical results shown in Figs.~\ref{fig:SP} and \ref{fig:LC}. We also confirmed that thermal and non-thermal emission in this model is consistent with the late-time observations at the optical and X-ray bands, respectively~\citep{BCM+18}.

\subsection{Radio--gamma-ray connection}
ALMA and NOEMA observations imply that the magnetization parameter may be  significantly different from that observed in the Crab nebula. 
Here we point out that high- and low-magnetization scenarios can be tested not only by high-frequency radio observations but also by soft \gray\ observations.  
As argued in \cite{MTO14}, sub-GeV \gRays\ can escape as early as optical photons. 
The Bethe-Heitler process is relevant for sub-GeV gamma rays, and its optical depth is $\tau_{\rm BH}\approx(8\sigma_{\rm BH}/\sigma_T)\tau_T\sim10\alpha_{\rm em}\tau_T$ for the CO ejecta, where $\tau_T$ is the Thomson optical depth and $\alpha_{\rm em}\simeq1/137$. 
Then, the gamma-ray breakout time is~\citep{MKK+15} 
\begin{eqnarray}
t_{\gamma-\rm bo}=t_{\rm pk}\sqrt{\frac{K_\gamma}{K\beta_{\rm ej}}}\sim t_{\rm pk}K_{\gamma,-2}^{1/2}K_{-1}^{-1/2}\beta_{\rm ej,-1}^{-1/2},
\end{eqnarray}
where $K_\gamma\sim0.01~{\rm g}^{-1}~{\rm cm}^{2}$ is the opacity at 0.1~GeV. 

\begin{figure}
\includegraphics[width=\linewidth]{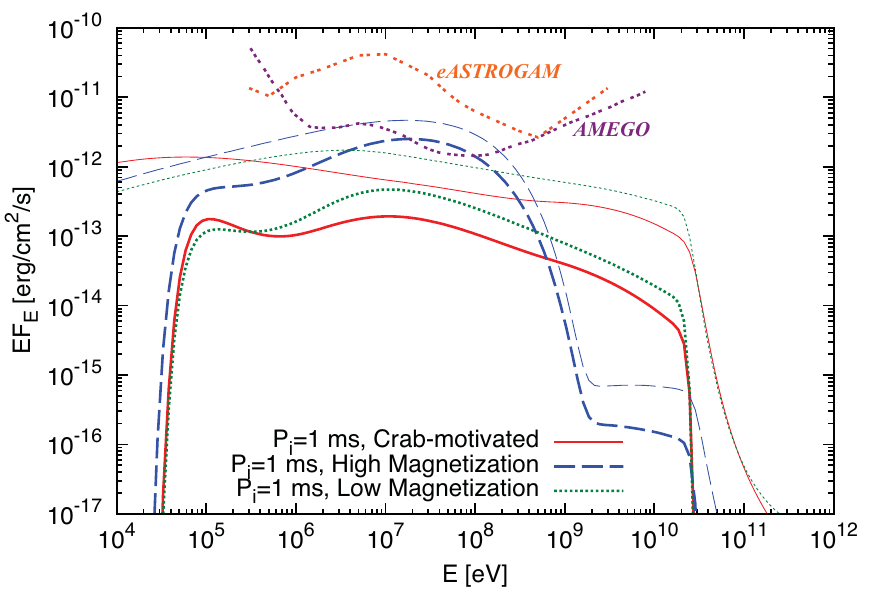}
\caption{Gamma-ray fluxes from SN 2017egm at $T_{\rm pk}$. Cases with/without gamma-ray attenuation in the SN ejecta are shown by thick (thin) curves. Sensitivities of eASTROGAM and AMEGO are shown for an integration time of $10^6$~s. 
}
\label{fig:gamma}
\end{figure}

In the high magnetization model, the peak energy of the intrinsic nebular spectrum, $E_b=h\nu_b$, is close to the synchrotron cutoff at $E_M\equiv h\nu_M\sim160$~MeV, which may be detected by gamma-ray telescopes sensitive to sub-GeV \gRays. 
As an example, the result for SN\,2017egm is shown in Fig.~\ref{fig:gamma}. 
The expected gamma-ray signal is difficult to detect with {\it Fermi}-LAT with its sensitivity of $\sim10^{-11}~{\rm erg}~{\rm cm}^{-2}~{\rm s}^{-1}$, but future MeV \gray\ telescopes such as AMEGO and eASTROGAM will have the sensitivity to probe the predicted fluxes around the \gray\ breakout time. 
The predictions are different from the lower-magnetization models, in which the inverse-Compton mechanism leads to gamma-rays beyond GeV energies~\citep{MKK+15,MTF+18}.  

%
%

The high-magnetization model also has a simple prediction for the relationship between radio and \gray\ fluxes,  
\begin{eqnarray}
\frac{\nu F_{\nu}|_{\rm soft\gamma}}{\nu F_{\nu}|_{\rm radio}}\approx{(2.42\times10^9)}^{2-\beta}
{\left(\frac{E_\gamma}{1~\rm MeV}\right)}^{2-\beta}{\left(\frac{\nu}{100~\rm GHz}\right)}^{\beta-2},
\end{eqnarray}
which leads to $(\nu F_{\nu}|_{\rm soft\gamma})/(\nu F_{\nu}|_{\rm radio})\sim5\times10^4$ for $\beta=1/2$ and $E_\gamma=h\nu <h\nu_b$. 
This is in contrast to the other two scenarios, in which the GeV \gRays\ are more prominent. 
Note that these results may be modified by attenuation in the ejecta (see Fig.~\ref{fig:gamma}). 


\section{Summary and discussion} 
The pulsar/magnetar-driven scenario for SLSNe-I naturally predicts synchrotron emission from embryonic PWNe.
Here we have reported ALMA and NOEMA observations of three SLSNe, SN\,2015bn, SN\,2016ard, and SN\,2017egm, at ages of 1--3~yr, which set new upper limits on their high-frequency radio emissions.   
We also presented details of a theoretical model to calculate nebular emission that can be approximated by the analytical prescription. 
In particular, for SN\,2015bn, we found that the upper limit at Band~6 ($\sim230$~GHz) disfavors the model if the nebular magnetization and lepton spectrum are similar to those for the Crab nebula. 
On the other hand, for SN\,2016ard and SN\,2017egm, the Crab-motivated model is still consistent with the ALMA and NOEMA limits, respectively, unless $P\sim P_{\rm max}$. 

The present millimetre limits~\citep[see also][]{LOK+19,EMO+21} are not sufficient to cover the relevant parameter space allowed by optical data, so further observations at the mm band are necessary to critically test the pulsar/magnetar-driven scenario. 
Dust emission observed at higher frequencies also enable us to probe the pulsar/magnetar central engine~\citep{OKM19}.   
Nevertheless, our results demonstrate that high-frequency radio emission provides a powerful probe for non-thermal activity associated with young SLSNe. 
We investigated alternative high- and low-magnetization models. 
In either case, the predicted synchrotron flux is well below the current upper limits, while avoiding the missing energy problem in SN\,2015bn~\citep{BCM+18}. 
In particular, the high magnetization model leads to prominent synchrotron emission in the $1-100$~MeV range, which are good targets for future 
\gray\ observatories such as AMEGO~\citep{Moiseev:2017mxg} and eASTROGAM~\citep{eASTROGAM17}.  

We note that high-frequency radio emission can also be produced in other scenarios for SLSNe. 
SLSNe may be accompanied by jets, and the resulting off-axis jet emission may produce variable radio emission~\citep{EMO+21,HTM+21}. 
An alternative mechanism is the interaction-powered scenario, in which particles are accelerated at the shocks between the SN ejecta and dense circumstellar material~\citep{SBN+16}. 
Synchrotron signatures of secondary electrons and positrons produced via $pp$ interactions are expected~\citep{MTO14}. 
Such a late interaction has been seen for Type Ibc SNe, and it is promising even for SLSNe-I.  

Fast-cooling nebular emission has been of interest as a counterpart signal of FRB sources and progenitors. 
The nascent nebular spectrum is predicted to have a steep spectrum with $\beta\geq1.5$~\citep{MKM16} and $\nu F_{\nu}|_{\rm radio}/(\nu F_{\nu}|_{\rm X-ray})\gtrsim8\times{10}^{-3}$. 
On the other hand, studies on Galactic PWNe infer $\beta\lesssim1.4$~\citep{GS06}, consistent with the slow cooling spectrum. 
It is important to pursue a flexible approach that does not assume a Crab-like spectrum in interpreting non-detections of optical and \xray\ counterparts.
The quasi-steady synchrotron flux is sensitive to not only $B_{*}$ and $P_i$ but also the age $t=T-T_i$. 
Slowly rotating magnetars, which are likely more common, are expected to yield lower radio fluxes. 
Thus, our model anticipates detectable persistent radio counterparts only for a fraction of FRBs like FRB\,121102.  


\section*{Acknowledgements}

The ALMA observations were performed based on the Cycle 5 ALMA proposal with the project code 2017.1.00975.S, titled with ``Searching for the Smoking Gun of Magnetar-Powered Super-Luminous Supernovae" (PI: Kohta Murase) and the Cycle 6 ALMA proposal with the project code 2018.1.01295.S, titled with ``A direct test of the magnetar-model in Superluminous Supernovae" (PI: Deanne Coppejans). 
ALMA is a partnership of ESO (representing its member states), NSF (USA) and NINS (Japan), together with NRC (Canada), MOST and ASIAA (Taiwan), and KASI (Republic of Korea), in cooperation with the Republic of Chile. The Joint ALMA Observatory is operated by ESO, AUI/NRAO and NAOJ. The National Radio Astronomy Observatory is a facility of the National Science Foundation operated under cooperative agreement by Associated Universities, Inc. The NOEMA observations were performed based on the Summer 2018 NOEMA proposal with the project ID S18BH, titled ``Testing the Magnetar-Powered Scenario for Super-Luminous Supernovae with NOEMA" (PI: Conor Omand). 
The work of K.M. is supported by the Alfred P. Sloan Foundation, NSF Grant No.~AST-1908689, and KAKENHI No.~20H01901 and No.~20H05852. 
C.M.B.O. has been supported by the Grant-in-aid for the Japan Society for the Promotion of Science (18J21778). 
The authors would like to thank Michel Bremer and Jan Martin Winters for their help with NOEMA data reduction.



\bibliographystyle{mnras}
\bibliography{ref}

\bsp	
\label{lastpage}

\end{document}